\documentclass[letterpaper, 10 pt, conference]{ieeeconf}
\usepackage{inputenc,amsmath,graphicx,amssymb}
\usepackage[space]{cite}
\usepackage{graphicx}
\usepackage{amsmath,bm,amsfonts}
\usepackage[version=4]{mhchem}
\usepackage{siunitx}
\usepackage{longtable,tabularx}
\usepackage{caption,subcaption,float}
\newcommand{\rom}[1]{\MakeUppercase{\romannumeral #1}}

\title{Decentralized Aerial Transportation and Manipulation of a Cable-Slung Payload With Swarm of Agents}

\author{\authorblockN{Aniket Sharma}
\authorblockA{Aerospace Engineering Department\\
Indian Institute of Technology, Madras\\
Chennai, Tamil Nadu 600036\\
Email: ae19s004@smail.iitm.ac.in}
\and
\authorblockN{Nandan K Sinha}
\authorblockA{Aerospace Engineering Department\\
Indian Institute of Technology, Madras\\
Chennai, Tamil Nadu 600036\\
Email: nandan@ae.iitm.ac.in}
}

\begin{document}

\maketitle

\begin{abstract}
With the advent of Unmanned Aerial Vehicles (UAV) and Micro Aerial Vehicles (MAV) in commercial sectors, their application for transporting and manipulating payloads has attracted many research works. A swarm of agents, cooperatively working to transport and manipulate a payload can overcome the physical limitations of a single agent, adding redundancy and tolerance against failures. In this paper, dynamics of a swarm connected to a payload \textit{via} flexible cables is modelled and a decentralized control is designed using Artificial Potential Field (APF). The swarm is able to transport the payload through an unknown environment to a goal position while avoiding obstacles from the local information received from the on-board sensors. The key contributions are (a) the cables are  modelled more accurately using lumped mass model instead of geometric constraints, (b) a decentralized swarm control is designed using potential field approach to ensure hover stability of system without payload state information, (c) the manipulation of payload elevation and azimuth angles are controlled by APF, and (d) the trajectory of the payload  for transportation is governed by potential fields generated by goal point and obstacles. The efficacy of the method proposed in this work are evaluated through numerical simulations under influence of external disturbances and failure of agents.
\end{abstract}

\section{Introduction}
The immense potential of Unmanned Aerial Vehicle and Micro-Aerial Vehicle autonomy presents possibilities of solutions to many fields such as exploration, surveillance, mapping, logistics and search and rescue \cite{Chung2018}.  Replacing humans from these mission with automated agents offers a faster and safer alternative. Drone delivery systems are gaining popularity because of the more convenient, versatile and faster solutions compared to ground based systems in congested areas \cite{Li2021}. In a post disaster scenario, UAVs and MAVs can be used for aerial transportation of supplies and establishing communications. To increase the payload capacity, a larger vehicle can be used or a number of agents can cooperatively work together. Introducing a swarm of agents for transporting a heavy payload adds flexibility and resilience to failure. A coordinated swarm can also manipulate the payload's orientation  by using the principles of a parallel manipulator \cite{Fink2011}. The complexity of a multi-agent system is increased from a single agent system but this paper proposes that the complexity can be reduced by studying the system as a swarm with decentralized control strategy. It should be possible to transport and manipulate a payload of any practical scale by increasing the size of swarm since the \textit{emergent} behaviour of a swarm is responsible for the task on a macro scale which depends only on local interaction of a single agent and does not affect the complexity of control law for an agent irrespective of the size of swarm.

Aerial transportation using cooperative multiple aerial manipulators has been studied in several research works. A survey on aerial transportation using single and multiple aerial UAVs is presented in \cite{Mohiuddin2020}. These research works include variations in the system based on the type of aerial platforms, manipulators and control of manipulator. The problem with scaling multi-rotor vehicles to accommodate for higher capacity of payload is that the thrust produced by rotors is increased by the square of the sweeping area of the rotors, which in turn requires increase in the size of the quadrotor. The volume, and hence the weight, of the multi-rotor increase by cube law; limiting the rotorcraft to a certain size and lowering its agility \cite{Mohiuddin2020}. When three or more multi-rotors lift the same payload, the slung load has less swinging during acceleration \cite{Klausen2020, dhiman2020autonomous}. Aerial manipulators can be largely categorized into 1) Link-based serial manipulator, 2) Parallel linked mechanisms, 3) Hydraulic manipulators and 4) Cable based manipulators. While the dexterity of cable based manipulators is none, the weight and changes in c.g. and inertial parameters is low \cite{Mohiuddin2020}. It is economically best and less prone to failures compared to other mechanisms and reduces constraints on the system. Cooperative hexacopters with  an n-DOF robotic arm are controlled to manipulate a payload of unknown mass and inertia by using online estimation of payload mass and inertia in \cite{Lee2017}. A team of omnidirectional quadrotor cooperatively estimate unknown payload parameters and manipulate the payload attached to the rotor frame \textit{via} grasping tool in \cite{Pierri2020}. The omnidirectional platform allows the vehicle to move in all directions without changing the payload attitude. Cable-based aerial transportation and manipulation has been studied in detail in several research works. A mathematical model for aerial manipulation that captures kinematic constraints and mechanics of underlying stable equilibria is presented in \cite{Michael2011}. Study of configurations of aerial robots that can admit multiple payload equilibrium positions by developing constraints on robotic configuration to address a unique payload position is presented in \cite{Fink2011}. The required tension in the cable is calculated from the available wrench set, payload dynamics and collision avoidance between UAVs and a decentralized output feedback control on fixed-time Extended State Observer is developed for a group of UAVs in \cite{Liu2021}. A leaderless distributed adaptive control algorithm was developed for aerial cooperative transportation in \cite{Arab2021} where cable forces and external disturbances are compensated using adaptive terms. The role of the internal force for stability of a beam-like payload is studied in \cite{Tognon2018} while manipulating it with two MAVs in a cooperative decentralized way. The cable is modeled as a massless constant length kinematic constraint in these works. A single quadrotor with flexible cable-slung load is studied in \cite{Lee2014}. Another study on cooperative transport of suspended load by two quadrotors with massless and stretchable cables modeled as high stiffness spring with a damping coefficient is presented in \cite{pizetta2016cooperative}. The deformation of the cables cannot be observed with geometric or massless spring model. A flexible cable model is used in this paper to study a more accurate representation of the payload swings. The cable model consists of a series of lumped masses of cable elements connected \textit{via} spring-damper systems. Monocular vision and inertial sensing is used in \cite{Li2021} for estimating the payload position to cooperatively control a cable suspended payload by a team of MAVs. Force/torque sensors on cables are used to estimate payload position in \cite{bernard2011autonomous}. Since the swarms generally consist of large number of agents, it is imperative for each agent to be economically as convenient as possible. The decentralized swarm formation proposed in this paper is independent of the payload state information. An inner PID (proportional-integral-derivative) control loop and an outer formation control loop is designed such that the formation is maintained in the presence of external disturbances, unequal mass distribution of payload, and failure of fellow agents. A decentralized swarm system introduces scalability and improves the flexibility of aerial transportation and manipulation systems. Dhiman et al. \cite{dhiman2020autonomous} proposed a control approach which uses a combination of PID and PD (proportional-derivative) controller for position control and transportation of the load by generating a formation trajectory of each vehicle based on an optimal trajectory of the load. Such centralized planning might not be available for pre-planning the flight mission in presence of an unknown obstacle-ridden environment. Various approaches for swarm formation control has been discussed in detail in \cite{Chung2018}. A behaviour based subsumption architecture is proposed in \cite{huang2021decentralised} for decentralized adaptive and energy efficient transportation of unknown load using swarm of quadrotors and analysis of experiments carried out in physics informed robot simulation platform is presented. Artificial potential field method was first introduced in \cite{khatib1985real}. An attractive potential drives the swarm agents to the swarm formation position while the repulsive potential ensures collision avoidance and equidistant spread in the formation. Bennet and McInnes proposed the use of bifurcating potential fields to alter swarm formation shapes by changing the bifurcation parameter of the potential fields \cite{Bennet2009}. A bounded artificial force in the form of exponential functions is presented in \cite{Qin2013} to formulate a control algorithm to navigate a swarm in a predefiend 2D shape while avoiding intermittent collisions. This paper proposes an artificial potential field based swarm formation control law, to navigate the agents connected to a payload \textit{via} cables in a formation and through an unknown environment while avoiding unknown obstacles to a goal set point. A novel manipulation control of the payload orientation is proposed by adjusting the attractive potential field of the swarm formation.

The rest of the paper is organized as follows: Section \rom{2} describes the problem and formulates it mathematically. Section \rom{3} models the dynamics of the payload, the swarm agents and the cable elements. Section \rom{4} proposes the controller for the gravity compensation, swarm formation, payload transportation and payload manipulation. The assumptions for the control law have been stated in this section and a hover stability analysis is presented for the payload. Section \rom{5} presents the results from numerical simulation of the system under different scenarios (obstacle avoidance, wind disturbance, and failure of an agent). Section \rom{6} ends the paper with the findings and the conclusion of the research.

\section{Problem formulation}

\begin{figure}[H]
    \centering
    \includegraphics[width=2.5in]{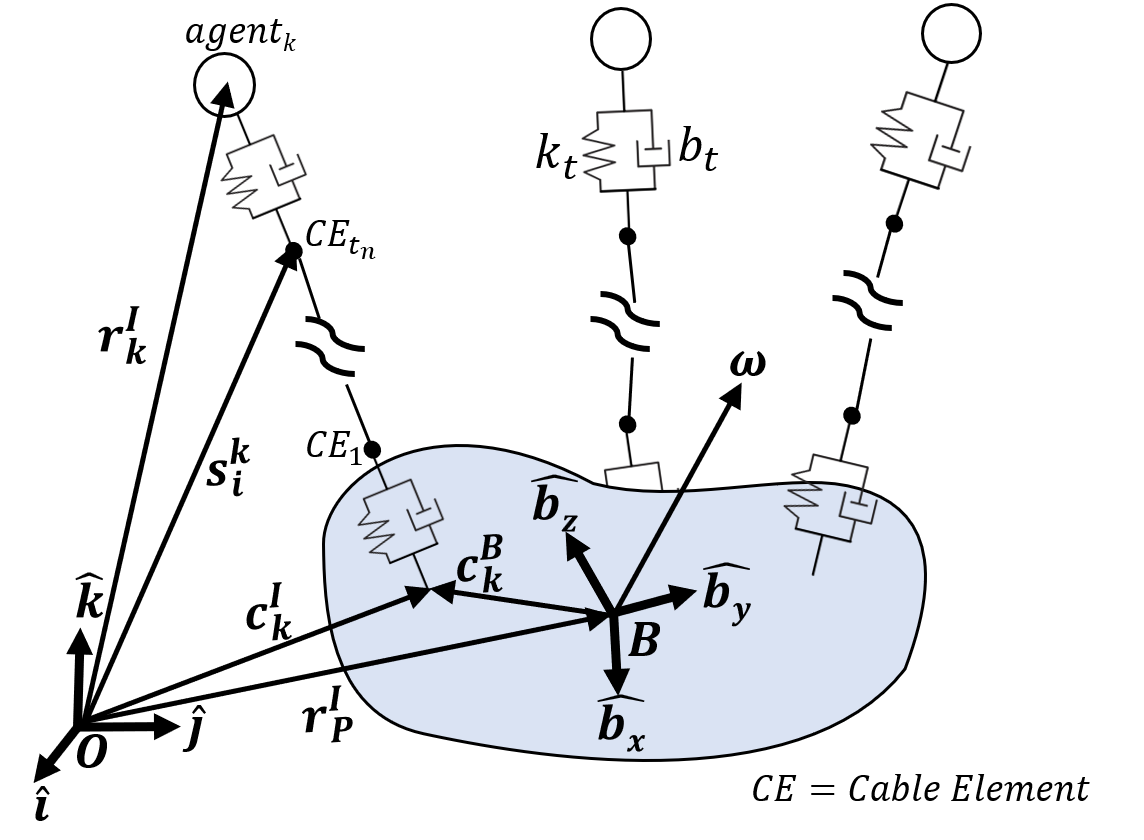}
    \caption{Schematic of swarm and payload system}
    \label{fig:1}
\end{figure}

Consider a rigid-body payload with center of mass at $B$ as shown in Fig \ref{fig:1}. $\mathcal{F_I}$ is the inertial frame of reference. The position vector of the payload's center of mass $B$ w.r.t. $\mathcal{F_I}$ is $\bm{r^I_P}$. There are $n$ swarm agents connected to the payload \textit{via} flexible cables where position vector of $i^{th}$ agent w.r.t. $\mathcal{F_I}$ is $\bm{r^I_i}$. $\bm{\hat{b}_x}$,$\bm{\hat{b}_y}$ and $\bm{\hat{b}_z}$ are the body-fixed frame of reference, $\mathcal{F_B}$, of the payload along the principal axes of the payload as shown in Fig \ref{fig:1} with the origin $B$ at center of mass of the payload.  
The interaction between payload and agents \textit{via} cables can be modelled as a series of lumped spring-mass-damper systems. The superscript of vectors describes the frame of reference, $I$ for the inertial reference frame and $B$ for the payload body-fixed reference frame.
\subsection{Relation between inertial and Body-fixed frame of references}
The transformation matrix from $\mathcal{F_B}$ to $\mathcal{F_I}$ is given by,
\begin{equation}
    \bm{^IT^B}=
    \begin{bmatrix}
    \bm{\hat{b}_x}&\bm{\hat{b}_y}&\bm{\hat{b}_z}\\
    \end{bmatrix}
    \label{eq:27}
\end{equation}
Since Eq. \eqref{eq:27} is an orthogonal matrix, inverse exists and the inverse is the transformation matrix from $\mathcal{F_I}$ to $\mathcal{F_B}$,
\begin{equation}
    \bm{^BT^I}=(\bm{^IT^B})^{-1}
    \label{eq:28}
\end{equation}
Position vector of $i^{th}$ agent in $\mathcal{F_B}$ is
\begin{equation}
    \implies \bm{r^B_i} = \bm{{}^BT^I}(\bm{r^I_i}-\bm{r^I_P})
    \label{eq:12}
\end{equation}

\section{Dynamics of the system}
The payload is considered as a rigid body of known mass. Under the assumption of a low-level attitude controller which controls the roll, pitch, yaw and thrust for the MAVs/UAVs, the dynamics of the multi-rotor can be modelled as point masses \cite{Klausen2020}. The forces experienced by both the swarm agents and the payload are Earth's gravity and tension from the cables. In addition, virtual control forces are generated by artificial potential fields to maintain swarm formation \cite{Bennet2009} and for transportation and maneuvering which is translated to desired roll, pitch, yaw and thrust inputs for the low-level controller.
\subsection{Cable model}
Let the cable be split into $t_n$ cable mass elements, where element $1$ is attached to the payload and element $t_n$ is attached to the agent. The position of cable element $i$ of agent $k$ is given by $\bm{s^k_i}$ w.r.t. $\mathcal{F_I}$ with mass $m_t$. Consider an agent $k$ with position vector $\bm{r^I_k}$ attached to the payload with position vector $\bm{r^I_P}$. The angular velocity vector of the payload is $\bm{\omega}$ and from Eq. \eqref{eq:27}, $\bm{^IT^B}$ is the transformation matrix from $\mathcal{F_B}$ to $\mathcal{F_I}$. The position vector of the cable anchor point with the payload and first cable element of agent $k$ is a constant vector in $\mathcal{F_B}$ frame, $\bm{c^B_k}$. $\bm{g} = [0,0,9.8]^T$ is the gravitational acceleration vector. Let $\bm{f}$ be a function defined as,
\begin{multline}
    \bm{f}(\bm{x_i},\bm{x_j},\bm{v_i},\bm{v_j})=-\alpha(\Delta x_{ij})\times\\
    \left(k_t\Delta x_{ij}+b_t\frac{\bm{x_{ij}}\cdot\bm{v_{ij}}}{\|\bm{x_{ij}}\|}\right) \frac{\bm{x_{ij}}}{\|\bm{x_{ij}}\|}
    \label{eq:29}    
\end{multline}

where $\alpha(x)=\left\{\begin{array}{ll}
       1, & \text{for } x>0\\
       0, & \text{otherwise}
       \end{array}\right\}$ is a scalar function which models slacking of the cable, $\bm{x_{ij} = \bm{x_i}-\bm{x_j}}$, $\bm{v_{ij} = \bm{v_i}-\bm{v_j}}$, $\Delta x_{ij}=\|\bm{x_{ij}}\|-l_{free}$ and $l_{free}$ is the nominal length of a single cable element.
Equation \eqref{eq:29} gives the resultant force on the cable element $i$ due to spring and damper action between the cable elements $i$ and $j$.

Using Eq. \eqref{eq:29}, for cable element $i = 2,3,...,t_{n}-1$,
\begin{multline}
    m_t \bm{\ddot{s}^k_i} = \bm{f}(\bm{s^k_i},\bm{s^k_{i+1}},\bm{\dot{s}^k_i},\bm{\dot{s}^k_{i+1}})+\bm{f}(\bm{s^k_i},\bm{s^k_{i-1}},\bm{\dot{s}^k_i},\bm{\dot{s}^k_{i-1}})\\
    -m_t\bm{g}
\label{eq:30}
\end{multline}

Using Eq. \eqref{eq:29}, for cable element $i = t_n$
\begin{multline}
    m_t \bm{\ddot{s}^k_i} = \bm{f}(\bm{s^k_i},\bm{r^I_k},\bm{\dot{s}^k_i},\bm{\dot{r}^I_k})+\bm{f}(\bm{s^k_i},\bm{s^k_{i-1}},\bm{\dot{s}^k_i},\bm{\dot{s}^k_{i-1}})\\
    -m_t\bm{g}
\label{eq:31}
\end{multline}

For cable element $i = 1$, $\bm{c^B_k}$ is a constant position vector of the point where the cable is anchored with the payload in $\mathcal{F^B}$. As shown in Fig. \ref{fig:1},
\begin{equation}
    \bm{c^I_k} = \bm{r^I_P}+ \bm{{}^IT^B}\bm{c^B_k}
    \label{eq:32}
\end{equation}

Differentiating Eq. \eqref{eq:32} w.r.t. time,
\begin{equation}
    \bm{\dot{c}^I_k} = \bm{\dot{r}^I_P}+ \bm{\omega\times c^B_k}
\label{eq:33}
\end{equation}

Using Eq. \eqref{eq:29}, \eqref{eq:32} and \eqref{eq:33}, for cable element $i = 1$,
\begin{multline}
    m_t \bm{\ddot{s}^k_i} = \bm{f}(\bm{s^k_i},\bm{c^I_k},\bm{\dot{s}^k_i},\bm{\dot{c}^I_k})+\bm{f}(\bm{s^k_i},\bm{s^k_{i+1}},\bm{\dot{s}^k_i},\bm{\dot{s}^k_{i+1}})\\
    -m_t\bm{g}
\label{eq:34}
\end{multline}

%%%%%%%%%%%%%%%%%%%%%%%%%%%%%%%%%%%%%%%%%%%%%%%%%%%%%%%%%%%%
\subsection{Dynamics of payload}
Using Eq. \eqref{eq:29}, \eqref{eq:32} and \eqref{eq:33}, for payload,
\begin{equation}
    m_P\bm{\ddot{r}^I_P} = -c\bm{\dot{r}^I_P}+\sum_{k=1}^n \left(\bm{f}(\bm{c^I_k},\bm{s^k_1},\bm{\dot{c}^I_k},\bm{\dot{s}^k_1})\right) -m_P\bm{g}
    \label{eq:35}
\end{equation}
where $c$ is a dissipation coefficient due to drag.

\begin{equation}
    \bm{I_P \dot{\omega}} = -\bm{\omega}\times(\bm{I_P \omega})+\bm{M}
    \label{eq:36}
\end{equation}
where $\bm{M} = \sum_{k=1}^n\left(\bm{{}^IT^B}\bm{c^B_k}\times\bm{f}(\bm{c^I_k},\bm{s^k_1},\bm{\dot{c}^I_k},\bm{\dot{s}^k_1})\right)$ is the external moment on the payload about $B$ in $\mathcal{F^I}$.

As per definition, $\mathcal{F^B} = \{\bm{\hat{b}_x}, \bm{\hat{b}_y}, \bm{\hat{b}_z}\}$ is the body-fixed axes of the payload.
\begin{equation}
\frac{d\bm{\hat{b}_i}}{dt} = \bm{\omega} \times \bm{\hat{b}_i}
\label{eq:37}
\end{equation}
for $i = x,y,z$ gives new orientation of the payload and new transformation matrix between $\mathcal{F^B}$ and $\mathcal{F^I}$ frames using Eq. \eqref{eq:27}.

\subsection{Dynamics of swarm agent}
Let the swarm consist of $n$ homogeneous agents with mass $m$. For a swarm agent $k$,
\begin{equation}
m \bm{\ddot{r}^I_k} = -c\bm{\dot{r}^I_k}+ \bm{f}(\bm{r^I_k},\bm{s^k_{t_n}},\bm{\dot{r}^I_k},\bm{\dot{s}^k_{t_n}})+\bm{u_k}(\bm{r^I_k})-m\bm{g}
\label{eq:38}
\end{equation}
where $\bm{u_k}(\bm{r^I_k})$ is the control input for swarm agent $k$.

\section{Controller design}
\begin{figure}
    \centering
    \includegraphics[width = 2in]{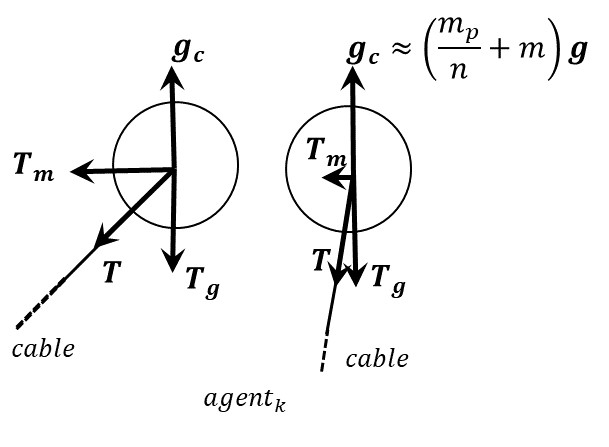}
    \caption{Compensation of gravitational pull}
    \label{fig:8}
\end{figure}

There can be different swarm formation configurations possible to carry out the transportation and manipulation task. A planar swarm formation such that the cables align parallel to the gravity vector allows the following:
\begin{itemize}
    \item Swarm \textit{x-y} formation is not affected by cable tensions at equilibrium conditions (Fig. \ref{fig:8}).
    \item As a result of independence from cable tension at equilibrium, the payload state information is not required for maintaining a proper formation structure.
    \item It also enables the independent design of \textit{x-y} formation (APF for transportation) and \textit{z} formation (APF for static equilibrium).
\end{itemize}
Further, the following assumptions are made about the system configuration:
\begin{enumerate}
    \item All the cables are of the same length and anchored at a common plane on the payload.
    \item The payload's center of mass is in the convex set of the anchor points.
    \item The size of swarm is greater than or equal to three agents.
    \item The swarm forms a connected graph i.e. no agent is isolated from the entire swarm.
    \item The swarm agents have the information of the size of the swarm
\end{enumerate}

For some time $t$, position of agent $k$ is $\bm{r^I_k}=[x_k,y_k,z_k]^T$. The total cable length is given by $L = l_{free}(t_n+1)$. Since the payload is connected physically with swarm agents \textit{via} cables, the attitude and position of the payload is also controlled by controlling swarm formation. Each agent has to compensate for its own weight and the tension of the cable due to payload's weight. 
As shown in Fig. \ref{fig:8}, a larger angle between the cable and the gravitational acceleration requires knowledge of cable tension vector to properly compensate the effects of the tension generated by the payload. The $T_m$ component of tension affects the swarm formation structure in \textit{xy} plane of the swarm formation. 

\subsection{Equilibrium of the system}
For ease of notation, let $\bm{f}(CE^*_i,CE^*_{i+1}) = \bm{f}(\bm{s^{k*}_i},\bm{s^{k*}_{i+1}},\bm{\dot{s}^{k*}_i},\bm{\dot{s}^{k*}_{i+1}})$. For $k^{th}$ cable, at equilibrium using Eq. \eqref{eq:30}, \eqref{eq:31}, and \eqref{eq:34},

\begin{equation}
    \bm{f}(CE^*_i,CE^*_{i+1})+\bm{f}(CE^*_i,CE^*_{i-1})-m_t\bm{g} =0
    \label{eq:100}
\end{equation}
\begin{equation}
    \bm{f}(CE^*_{t_n},\bm{r}^{I*}_k)+\bm{f}(CE^*_{t_n},CE^*_{t_{n-1}})-m_t\bm{g} =0
    \label{eq:101}
\end{equation}
\begin{equation}
    \bm{f}(CE^*_1,CE^*_{2})+\bm{f}(CE^*_1,c^{I*}_k)-m_t\bm{g} =0
    \label{eq:102}
\end{equation}
From Eq. \eqref{eq:102},
\begin{align}
    \bm{f}(CE^*_1,c^{I*}_k) &= m_t\bm{g}-\bm{f}(CE^*_1,CE^*_{2})\\
    &= m_t\bm{g}+\bm{f}(CE^*_2,CE^*_1)
    \label{eq:103}    
\end{align}
From Eq. \eqref{eq:100},
\begin{align}
    \bm{f}(CE^*_i,CE^*_{i-1}) &= m_t\bm{g}-\bm{f}(CE^*_i,CE^*_{i+1})\\
    &= m_t\bm{g}+\bm{f}(CE^*_{i+1},CE^*_i)
    \label{eq:104}    
\end{align}
Using Eq. \eqref{eq:104} in Eq. \eqref{eq:103} recursively,
\begin{equation}
\bm{f}(CE^*_1,c^{I*}_k) = t_n m_t\bm{g} + \bm{f}(r^{I*}_k,CE^*_{t_n})
    \label{eq:105}
\end{equation}
At equilibrium, from Eq. \eqref{eq:38},
\begin{equation}
\bm{f}(r^{I*}_k,CE^*_{t_n}) =  m\bm{g} - \bm{u_k(r^{I*}_k)}
    \label{eq:106}
\end{equation}
Using Eq. \eqref{eq:106} in Eq. \eqref{eq:105},
\begin{equation}
\bm{f}(CE^*_1,c^{I*}_k) = t_n m_t\bm{g} + m\bm{g} - \bm{u_k(r^{I*}_k)} 
    \label{eq:107}
\end{equation}
From Eq. \eqref{eq:35}, rigid-body translational dynamics equation at equilibrium is,
\begin{equation}
\sum_{k=1}^n \left(\bm{f}(c^{I*}_k,CE^*_1)\right) -m_P\bm{g} = 0
    \label{eq:108}
\end{equation}
Putting Eq. \eqref{eq:107} in \eqref{eq:108},
\begin{equation}
\sum_{k=1}^n \left(t_n m_t\bm{g} + m\bm{g} - \bm{u_k(r^{I*}_k)}\right) +m_P\bm{g} = 0
    \label{eq:109}
\end{equation}
Simplifying Eq. \eqref{eq:109}, 
\begin{equation}
\sum_{k=1}^n \bm{u_k(r^{I*}_k)}  = n(t_n m_t\bm{g} + m\bm{g}) + m_P\bm{g}
    \label{eq:109}
\end{equation}
From Eq. \eqref{eq:36}, rotational equilibrium is,
\begin{equation}
\sum_{k=1}^n\left(\bm{c^I_k}\times\bm{f}(c^{I*}_k,CE^*_1)\right) = 0
    \label{eq:110}
\end{equation}
Using Eq. \eqref{eq:107} in \eqref{eq:110},
\begin{equation}
\sum_{k=1}^n\left(\bm{c^I_k}\times(t_n m_t\bm{g} + m\bm{g} - \bm{u_k(r^{I*}_k)})\right) = 0
    \label{eq:111}
\end{equation}

Let control input be split into two terms such that,
\begin{equation}
    \bm{u_k(r^{I*}_k)} = \bm{u_g} +     \bm{u_f(r^{I*}_k)}
    \label{eq:112}
\end{equation}
where $\bm{u_g}$ is common gravity compensation term and $\bm{u_f(r^{I*}_k)}$ is the formation control term. Let $\bm{u_g} = (m + t_n m_t + \frac{m_P}{n})\bm{g}$, then simplifying Eq. \eqref{eq:109} and \eqref{eq:111},
\begin{equation}
    \sum_{k=1}^n \bm{u_f(r^{I*}_k)}  = 0
    \label{eq:113}
\end{equation}
\begin{equation}
    \sum_{k=1}^n\bm{c^I_k}\times  \bm{u_f(r^{I*}_k)}) = -\sum_{k=1}^n\bm{c^I_k}\times\frac{m_P}{n}\bm{g}    
    \label{eq:114}
\end{equation}

Simplifying the swarm and payload equation in Eq. \eqref{eq:29}, \eqref{eq:35} and \eqref{eq:38} as,

\begin{multline}
m \bm{\ddot{r}^I_k} = -c\bm{\dot{r}^I_k}-\left(\frac{k_t}{t_n+1}(\|\bm{d^I_k}\|-L)+\frac{b_t}{t_n+1}\frac{\bm{d^I_k}\cdot\bm{\dot{d}^I_k}}{\|\bm{d^I_k}\|}\right) \\
\times \frac{\bm{d^I_k}}{\|\bm{d^I_k}\|} +\bm{u_g}(\bm{r^I_k})-m\bm{g}
\label{eq:52}    
\end{multline}

where $\bm{u_g}(\bm{r^I_k})$ is the gravity compensation input for $k^{th}$ agent. 
\begin{multline}
m_P \bm{\ddot{r}^I_P} = -c\bm{\dot{r}^I_P}+\sum_{k=1}^n\left(\left(\frac{k_t}{t_n+1}(\|\bm{d^I_k}\|-L)+\right.\right.\\
\left.\left.\frac{b_t}{t_n+1}\frac{\bm{d^I_k}\cdot\bm{\dot{d}^I_k}}{\|\bm{d^I_k}\|}\right) \times \frac{\bm{d^I_k}}{\|\bm{d^I_k}\|}\right)-m_P\bm{g}
\label{eq:53}    
\end{multline}

At equilibrium, $\bm{\ddot{r}^{I*}_k} = \bm{\dot{r}^{I*}_k} = 0$ for $k = 1,2,...n$ and $\bm{\ddot{r}^{I*}_P} = \bm{\dot{r}^{I*}_P} = 0$. Let $\bm{r^{I*}_k}-\bm{c^{I*}_k} = \bm{d^{I*}_k}$. Simplifying Eq. \eqref{eq:52} at equilibrium,

\begin{multline}
    \left(\frac{k_t}{t_n+1}(\|\bm{d^{I*}_k}\|-L)+\frac{b_t}{t_n+1}\frac{\bm{d^{I*}_k}\cdot\bm{\dot{d}^{I*}_k}}{\|\bm{d^{I*}_k}\|}\right)\\\times \frac{\bm{d^{I*}_k}}{\|\bm{d^{I*}_k}\|} = \bm{u_g}(\bm{r^{I*}_k})-m\bm{g}
\label{eq:54}  
\end{multline}

Using Eq. \eqref{eq:54} in Eq. \eqref{eq:53},
\begin{align}
\sum_{k=1}^n\left(\bm{u_g}(\bm{r^{I*}_k})-m\bm{g}\right)-m_P\bm{g} &= 0\nonumber\\
\implies\sum_{k=1}^n\left(\bm{u_g}(\bm{r^{I*}_k})\right) &= (m_P+nm)\bm{g}
\label{eq:55}    
\end{align}
Using Eq. \eqref{eq:55}, a valid control for gravity compensation can be defined as,
\begin{equation}
\bm{u_g}(\bm{r^{I*}_k}) = \left(\frac{m_P}{n}+m\right)\bm{g}
\label{eq:56}    
\end{equation}
Substituting Eq. \eqref{eq:56} in Eq. \eqref{eq:54},

\begin{multline}
    \left(\frac{k_t}{t_n+1}(\|\bm{d^{I*}_k}\|-L)+\frac{b_t}{t_n+1}\frac{\bm{d^{I*}_k}\cdot\bm{\dot{d}^{I*}_k}}{\|\bm{d^{I*}_k}\|}\right) \\ \times \frac{\bm{d^{I*}_k}}{\|\bm{d^{I*}_k}\|} = \frac{m_P}{n}\bm{g}
\label{eq:57}  
\end{multline}

Equation \eqref{eq:57} shows that for the gravity compensation designed in Eq. \eqref{eq:56}, the tension vector, and hence the cable, is aligned with the gravitational acceleration vector at equilibrium.
\subsection{Stability analysis}
Let $\mathcal{F^A}$ be a reference frame parallel to $\mathcal{F^I}$ and origin at anchor point of $k^{th}$ agent, $\bm{c^I_k}$. A small disturbance $\bm{\delta r^A_k}$ to the position vector of $k^{th}$ agent, $\bm{r^A_k}=\bm{r^{A*}_k}+\bm{\delta r^A_k}$ is introduced. For small disturbances, the deformation in the cable can be neglected, implying the length of the disturbed cable vector is same. Using Eq. \eqref{eq:52} and \eqref{eq:56},
\begin{multline}
    m \bm{\ddot{r}^{A}_k} = -c\bm{\dot{r}^{A}_k}-\left(\frac{k_t}{t_n+1}(\|\bm{r^{A}_k}\|-L)+\frac{b_t}{t_n+1}\frac{\bm{r^{A}_k}\cdot\bm{\dot{r}^{A}_k}}{\|\bm{r^{A}_k}\|}\right) \\ 
    \times \frac{\bm{r^{A}_k}}{\|\bm{r^{A}_k}\|} +\frac{m_P}{n}\bm{g}
    \label{eq:58}
\end{multline}
$\|\bm{r^{A*}_k}+\bm{\delta r^{A}_k}\| = \|\bm{r^{A*}_k}\| = \gamma$ for no deformation in cable due to disturbance for some $\gamma > L$. Since at equilibrium, cable tension vector is aligned with gravitational acceleration vector,  $\bm{r^{A*}_k} = (0,0,\gamma)^T$. 
\begin{align}
\begin{split}
    (\bm{r^{A*}_k}+\bm{\delta r^{A}_k})\cdot(\bm{\dot{r}^{A*}_k}+\bm{\dot{\delta r}^{A}_k}) ={}& \bm{r^{A*}_k}\cdot\bm{\dot{r}^{A*}_k}                     +\bm{r^{A*}_k}\cdot\bm{\dot{\delta                      r}^{A}_k}\\
               &  +\bm{\dot{r}^{A*}_k}\cdot\bm{\delta r^{A}_k}
\end{split}\nonumber\\
\begin{split}
 ={} & \bm{r^{A*}_k}\cdot\bm{\dot{\delta r}^{A}_k}
\label{eq:59}
\end{split}
\end{align}

Simplifying Eq. \eqref{eq:58} using Eq. \eqref{eq:59},
\begin{multline}
    m \bm{\delta\ddot{ r}^{A}_k} = -c\bm{\delta\dot{ r}^{A}_k}-\left(\frac{k_t}{t_n+1}(\|\bm{r^{A*}_k}\|-L) 
    \right.\\
    \left.+\frac{b_t}{t_n+1}\frac{\bm{r^{A*}_k}\cdot\bm{\dot{\delta r}^{A}_k}}{\|\bm{r^{A*}_k}\|}\right) \frac{\bm{\delta r^{A}_k}}{\|\bm{r^{A*}_k}\|}\\
    -\frac{b_t}{t_n+1}\left(\frac{\bm{r^{A*}_k}\cdot\bm{\dot{\delta r}^{A}_k}}{\|\bm{r^{A*}_k}\|}\right)\frac{\bm{r^{A*}_k}}{\|\bm{r^{A*}_k}\|}
    \label{eq:60}
\end{multline}

Reducing higher order terms to zero in Eq. \eqref{eq:60},

\begin{multline}
    m \bm{\delta\ddot{ r}^{A}_k} = -c\bm{\delta\dot{r}^{A}_k}-\frac{k_t(\gamma-L)}{\gamma(t_n+1)}\bm{\delta r^{A}_k}\\
    -\frac{b_t\bm{r^{A*}_k}\cdot\bm{\dot{\delta r}^{A}_k}}{\gamma^2(t_n+1)}\bm{r^{A*}_k}    
\end{multline}
\begin{multline}
\label{eq:61}
    m \begin{bmatrix}
    \delta \ddot{x}_k\\ \delta \ddot{y}_k\\ \delta \ddot{z}_k\\
    \end{bmatrix}
    = -c\begin{bmatrix}
    \delta \dot{x}_k\\ \delta \dot{y}_k\\ \delta \dot{z}_k\\
    \end{bmatrix}-\frac{k_t(\gamma-L)}{\gamma(t_n+1)}
    \begin{bmatrix}
    \delta x_k\\ \delta y_k\\ \delta z_k\\
    \end{bmatrix}
    \\
    -\frac{b_t\delta \dot{z}_k}{\gamma(t_n+1)}\begin{bmatrix}
    0\\ 0\\ \gamma\\
    \end{bmatrix}    
\end{multline}

\begin{equation}
    \begin{split}
    m \delta \ddot{x}_k &= -c\delta \dot{x}_k-\frac{k_t(\gamma-L)}{\gamma(t_n+1)}\delta x_k\\
    m \delta \ddot{y}_k &= -c\delta \dot{y}_k-\frac{k_t(\gamma-L)}{\gamma(t_n+1)}\delta y_k\\
    m \delta \ddot{z}_k &= -\left(c+\frac{b_t}{(t_n+1)}\right)\delta \dot{z}_k-\frac{k_t(\gamma-L)}{\gamma(t_n+1)}\delta z_k\\
\end{split}
\label{eq:62}
\end{equation}
The roots of the characteristic equation of the first two second order linear differential equations in Eq. \eqref{eq:62} are,
\begin{equation}
    \lambda_{x,y} = \frac{-\frac{c}{m}\pm\sqrt{\left(\frac{c}{m}\right)^2-4\frac{k_t(\gamma-L)}{m\gamma(t_n+1)}}}{2}
\end{equation}
The roots have negative real part in case of complex or repeating roots. For distinct real roots,
\begin{equation}
    \begin{split}
        \frac{k_t(\gamma-L)}{m\gamma(t_n+1)}&>0\\
        -4\frac{k_t(\gamma-L)}{m\gamma(t_n+1)}&<0\\
        \left(\frac{c}{m}\right)^2-4\frac{k_t(\gamma-L)}{m\gamma(t_n+1)}&<\left(\frac{c}{m}\right)^2\\
        -\frac{c}{m}<\sqrt{\left(\frac{c}{m}\right)^2-4\frac{k_t(\gamma-L)}{m\gamma(t_n+1)}}&<\frac{c}{m}\\
        -\frac{c}{m}+\sqrt{\left(\frac{c}{m}\right)^2-4\frac{k_t(\gamma-L)}{m\gamma(t_n+1)}}&<0
    \end{split}
    \label{eq:63}
\end{equation}
Equation \eqref{eq:63} shows that the real part of the root is always negative, implying a stable solution, i.e. the disturbance decays to zero in x-y dimensions.

The roots of the characteristic equation of the third second order linear differential equation in Eq. \eqref{eq:62} are,
\begin{multline}
    \lambda_{z} = \frac{1}{2}\times\left\{ -\frac{c}{m}-\frac{b_t}{m(t_n+1)}\pm\right.\\
    \left.\sqrt{\left(\frac{c}{m}+\frac{b_t}{m(t_n+1)}\right)^2-4\frac{k_t(\gamma-L)}{m\gamma(t_n+1)}}\right\}
\end{multline}
With the same steps as shown in Eq. \eqref{eq:63}, since $\frac{b_t}{m(t_n+1)}>0$, the real part of the root is always negative, implying a stable solution, i.e. the disturbance decays to zero in z dimension as well. This proves the linear stability of the equilibrium of cable vectors along gravitational acceleration vector.

%%%%%%%%%%%%%%%%%%%%%%%%%%%%%%%%%%%%%%%%%%%%%%%%%%%%%%%%
\subsection{Swarm formation}
\subsection{Gravity compensation}
\subsubsection{Altitude of the payload}
\begin{figure}
    \centering
    \includegraphics[width = 2in]{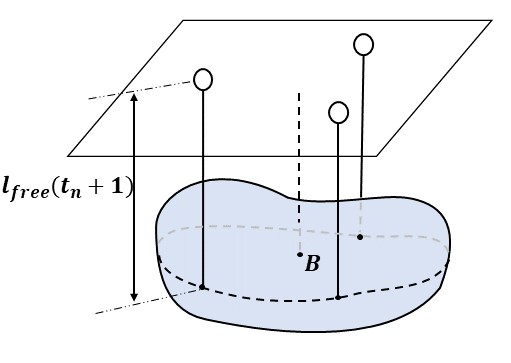}
    \caption{Swarm formation structure}
    \label{fig:10}
\end{figure}
Let $\bm{r^I_g}=[x_g,y_g,z_g]^T$ be some goal set point of the payload. The swarm tracks a center, $S$, with position vector $\bm{p}^I$. As shown in Fig. \ref{fig:9}, the azimuth and elevation angle of a normal to the plane of swarm formation and passing through $S$ w.r.t. $\mathcal{F^I}$ are $\psi$ and $\theta$ respectively. As shown in Fig. \ref{fig:10}, an $x-y$ plane at a height of $z_g+L$ is proposed as an attractor for the swarm. Using assumptions 1, 2 and 3, for equal cable lengths of each agent, the plane of swarm will be parallel to the plane of anchor points as the swarm tries to reach this attractor. This ensures orientation and z-axis stability of the payload. The choice of aligning cables with gravitational acceleration allows the decoupling of $x-y$ plane states from other states of the payload. 
For swarm agent $k$, an attractive artificial force magnitude in z-direction is defined as:
\begin{equation}
    f_z(\bm{r^I_k},\theta,\psi) = k_z\left(z_k-(z_g+L+\delta_k(\theta,\psi))\right)
    \label{eq:5}
\end{equation} where $k_z$ is a gain constant.

This field attracts the swarm agents on a plane at a height of $z_g+L+\delta_k$. This height ensures tension in the cables. $\delta_k$ controls the azimuth and elevation angles of the payload which will be discussed in the subsequent subsection.

\subsubsection{Manipulation of the payload}
\begin{figure}
    \centering
    \includegraphics[width=\columnwidth]{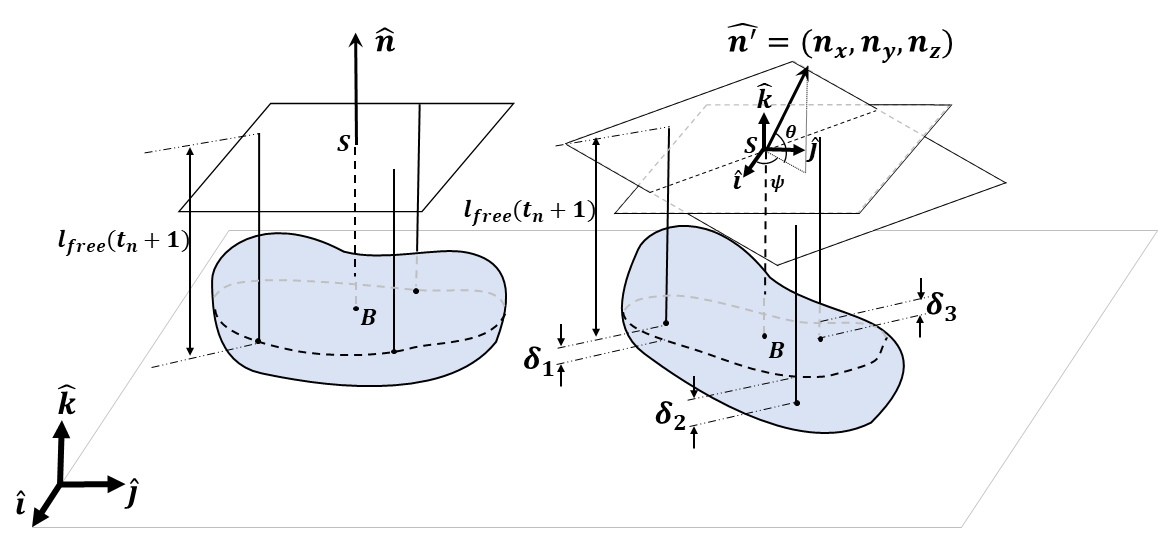}
    \caption{Manipulation of the payload}
    \label{fig:9}
\end{figure}
Figure \ref{fig:9} (left) shows a payload carried by a swarm consisting of three agents by maintaining a planar formation at a given height. A normal unit vector passing through $S$, $\bm{\hat{n}'} = [n_x,n_y,n_z]^T$, is defined by the desired azimuth ($\psi$) and elevation angle ($\theta$) of the payload as follows,
\begin{equation}
    \bm{\hat{n}'} = [\cos{\theta}\cos{\psi},\cos{\theta}\sin{\psi},\sin{\theta}]^T
    \label{eq:42}
\end{equation}
As shown in Fig.\ref{fig:9} (right), to reach the attitude of the payload defined by $\bm{\hat{n}'}$, the agents have to compensate for the increments/decrements, $\delta_k$, of the projections of the cable on the tilted plane defined by $\bm{\hat{n}'}$ and passing through $S$. The change $\delta_k$ for agent $k$ is given by,
\begin{equation}
    \delta_k = \frac{-n_x (x_k-p_x)-n_y(y_k-p_y)}{n_z}
    \label{eq:43}
\end{equation}
where $[p_x,p_y,p_z] = \bm{p^I}$.

\subsubsection{Transportation of the payload}
Let $\bm{\Delta^k}=\bm{r^I_k}-\bm{p^I}=[\Delta^k_x,\Delta^k_y,\Delta^k_z]^T$ and $\|\Delta^k_{xy}\|=\sqrt{{\Delta^k_x}^2+{\Delta^k_y}^2}$for $k^{th}$ agent. A bounded transportation force, $f_{x-y}(\bm{r^I_k},\bm{p^I}):\mathcal{R}^2\mapsto[0,1) $, is defined as follows,
\begin{equation}
    \label{eq:64}
    f_{x-y}(\bm{r^I_k},\bm{p^I}) = 1-\frac{(1+e^{\beta})^2}{(1+e^{-\|\Delta^k_{xy}\|+\beta})(1+e^{\|\Delta^k_{xy}\|+\beta})}
\end{equation}
\begin{figure}
    \centering
    \includegraphics[width=2in]{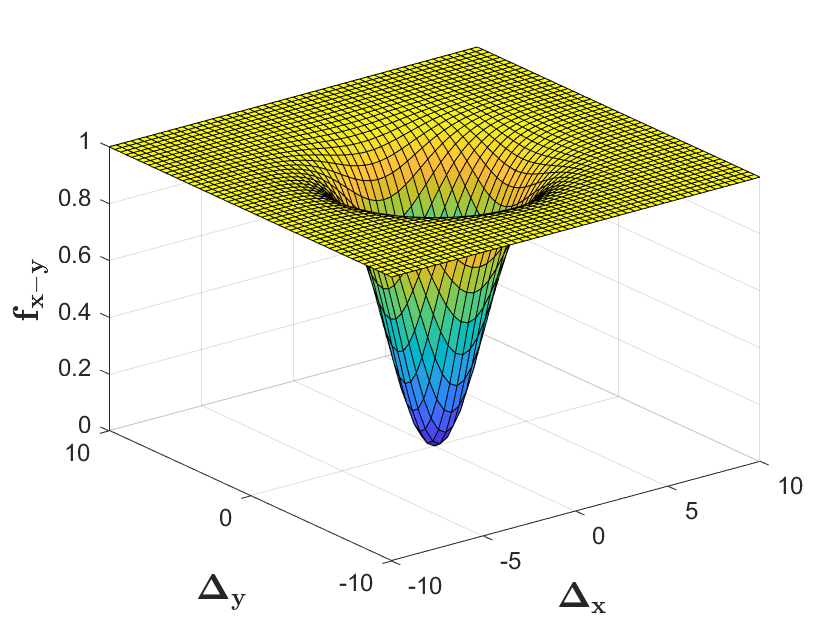}
    \caption{Transportation field}
    \label{fig:tfield}
\end{figure}
The nature of this field is shown in Fig. \ref{fig:tfield}. The force is bounded and constant; it decays exponentially as the distance from origin decreases. The parameter $\beta$ can be varied to vary the area of negligible force. This field allows the swarm agent's x-y position to lie on a circular disc centered at $\bm{p^I}$. With evolution of $\bm{p^I}$ with time, the swarm follows this attractor field in x-y plane. Irrespective of the contour or shape of the payload, with appropriate $\beta$ value, this attractor field can track the desired swarm center while maintaining the cables aligned with gravitational acceleration. Using Eq. \eqref{eq:5} and \eqref{eq:64}, the net attractive field on an agent $k$ of the swarm is,
\begin{equation}
    \label{eq:66}
    \bm{[f_A]^I_k}= \frac{f_{x-y}(\bm{r^I_k},\bm{p^I})}{\|\Delta^k_{xy}\|}\begin{bmatrix}
    \Delta^k_x \\ \Delta^k_y \\ 0
    \end{bmatrix} + f_{z}\begin{bmatrix}
    0 \\ 0 \\ 1
    \end{bmatrix}
\end{equation}

Let $\mathcal{N}_k$ be set of $\mathbb{R}^3$ position vectors of swarm agents in vicinity of $k^{th}$ agent. To avoid collision between agents and equidistant spacing amongst the agents in formation, a repulsive potential field is defined as \cite{Bennet2009},
\begin{equation}
    [U_R(\bm{r^I_k})]^I_k = \sum_{\bm{r^I_j}\in\mathcal{N}_k}C_R e^\frac{-\|\bm{r^I_k}-\bm{r^I_j}\|}{L_R}
    \label{eq:6}
\end{equation}

Let $\mathcal{O}_k$ be set of $\mathbb{R}^3$ position vectors of obstacles in vicinity of $k^{th}$ agent. A repulsive potential alters the goal maneuver of the swarm from Eq. \eqref{eq:64} to avoid obstacles.
\begin{equation}
    [U_O(\bm{r^I_k})]^I_k = \sum_{\bm{r^I_o}\in\mathcal{O}}C_o e^\frac{-\|\bm{r^I_k}-\bm{r^I_o}\|}{L_o}
    \label{eq:45}
\end{equation}

\subsubsection{Swarm center evolution}
For a successful manipulation using the offsets calculated in Eq. \eqref{eq:43}, the actual swarm center has to be close to $S$. A slower evolution of $\bm{p^I}$ in x-y plane can ensure close tracking of the actual swarm center; but the fast control response of azimuth and elevation angles for the payload is desirable. Hence, the evolution of swarm center for tracking is designed to respond faster in z-direction than in x-y plane. Let $\bm{\zeta}=\bm{r^I_g}-\bm{p^I}=[\zeta_x,\zeta_y,\zeta_z]^T$. $\bm{p^I}$ evolves according to the Ordinary Differential Equation,
\begin{equation}
    \label{eq:65}
    \bm{\dot{p}^I} = -\frac{(1-e^\frac{-\|\bm{\zeta}\|}{L_S})}{L_S}    \text{diag}([e^{-\|\zeta_z\|},e^{-\|\zeta_z\|},1])
    \bm{C_S}\frac{\bm{\zeta}}{\|\bm{\zeta}\|}
\end{equation}
where $\text{diag}([e^{-\|\zeta_z\|},e^{-\|\zeta_z\|},1])$ is a $3\times3$ diagonal gain matrix ensuring slower x-y response till z-axis response settles and $\bm{C_S}$ is a constant $3\times3$ diagonal gain matrix.

\subsubsection{Block diagram of the control scheme}
\begin{figure}
    \centering
    \includegraphics[width=\columnwidth]{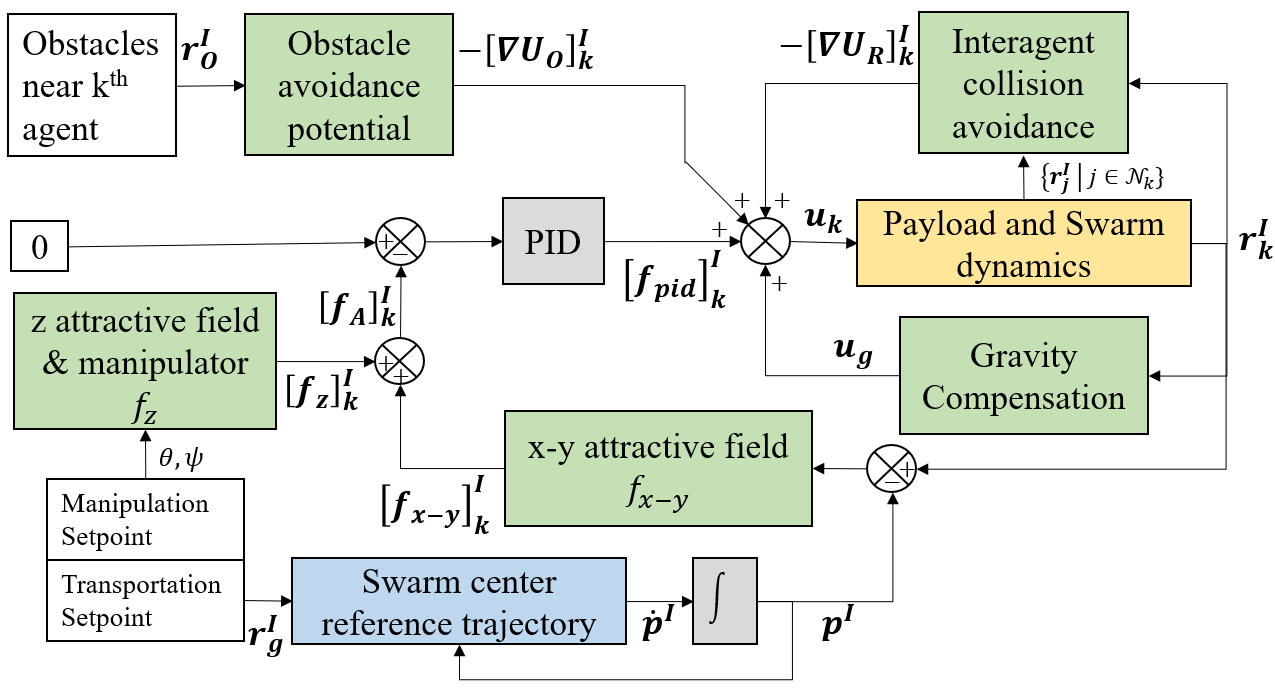}
    \caption{Block diagram of swarm formation control}
    \label{fig:pid}
\end{figure}

A generic configuration of the swarm anchor points on the payload can be asymmetric. As a result, there can be imbalanced moments on the payload unaccounted in the swarm formation laws discussed above. To mitigate these unaccounted disturbances and moments, the attractive field for every agent is passed through a PID (Proportional Integral Derivative) block. The overall swarm formation control block diagram is shown in Fig. \ref{fig:pid}. Both the fields in Eq. \eqref{eq:66} are monotonically increasing with global minima at origin, which is the desired equilibrium space for the formation. Thus, the setpoint of the PID block is origin for $\bm{[f_A]^I_k}$ as shown in Fig. \ref{fig:pid}. 

\subsubsection{Decentralized formation control}
Using Eq. (\ref{eq:56}-\ref{eq:65}), the control input of agent $k$ is given by,
\begin{multline}
    \bm{u_k}(\bm{r^I_k}) = \bm{u_g}(\bm{r^I_k})+\bm{[f_{pid}]^I_k}-\nabla[U_R(\bm{r^I_k})]^I_k\\
    -\nabla[U_O(\bm{r^I_k})]^I_k
    \label{eq:40}
\end{multline}

\section{Simulation results}
The setpoint for the payload is $[15,15,10]^T$ and the desired azimuth and elevation angle of the payload is $-60^{\circ}$ and $60^{\circ}$ respectively. The shape of the payload is considered to be a cylinder of radius $r_P=5m$ and $l_P = 10m$ . The inertia tensor of the payload is given by,
$$
    I_P = \begin{bmatrix}
    291.67 &0 &0\\0 &291.67 &0\\0 &0 &250
    \end{bmatrix}
\text{kg-m}^2
$$ 
The parameters for the simulation are tabulated in Table \ref{table:1}.
\begin{table}[H]
\renewcommand{\arraystretch}{1.3}
\caption{Simulation parameters}
\centering
\begin{tabular}{c||c}
\hline
\bfseries Parameter & Value\\
\hline\hline
 $m_P$ & 20kg\\
 $n$ & 7\\
 $m$ & 1.3kg\\
 $m_t$ & 0.003kg\\
 \hline
\end{tabular}
\begin{tabular}{c||c}
\hline
\bfseries Parameter & Value\\
\hline\hline
  $t_n$ & 2\\
 $l_{free}$ & 1.5m\\
 $k_t$ & 10073N/m\\
 $b_t$ & 0.1Ns/m\\
 \hline
\end{tabular}
\begin{tabular}{c||c}
\hline
\bfseries Parameter & Value\\
\hline\hline
 $C_{s}$ & diag($[2,2,20]$) \\
 $L_{s}$ & 5\\
 $\beta$ & 2\\
 $Kp$ & diag($[2,2,4]$)\\
 $Ki$ & diag($[0,0,0.5]$)\\
 \hline
\end{tabular}
\begin{tabular}{c||c}
\hline
\bfseries Parameter & Value\\
\hline\hline
 $Kd$ & diag($[0,0,8]$)\\
 $C_R$ & 0.1\\
 $L_R$ & 1\\
 $C_o$ & 500\\
 $L_o$ & 3\\
 \hline
\end{tabular}
\label{table:1}
\end{table}
The mission is simulated for three scenario using MATLAB ode45 solver. The performance of this decentralized aerial transportation and manipulation system is shown in the presence of an obstacle in Case 1. Case 2 studies the performance of this novel aerial transportation and manipulation system in presence of wind disturbances. Case 3 simulates a failure (loss of complete control) on one agent of the swarm. The swarm system is resilient to the unaccounted factors in the Cases 2 and 3, and successfully carries out the transportation and manipulation of payload without much degradation of performance.

\subsection{Case 1: Transportation and manipulation in presence of an obstacle}
\begin{figure}
    \centering
    \includegraphics[width=\columnwidth]{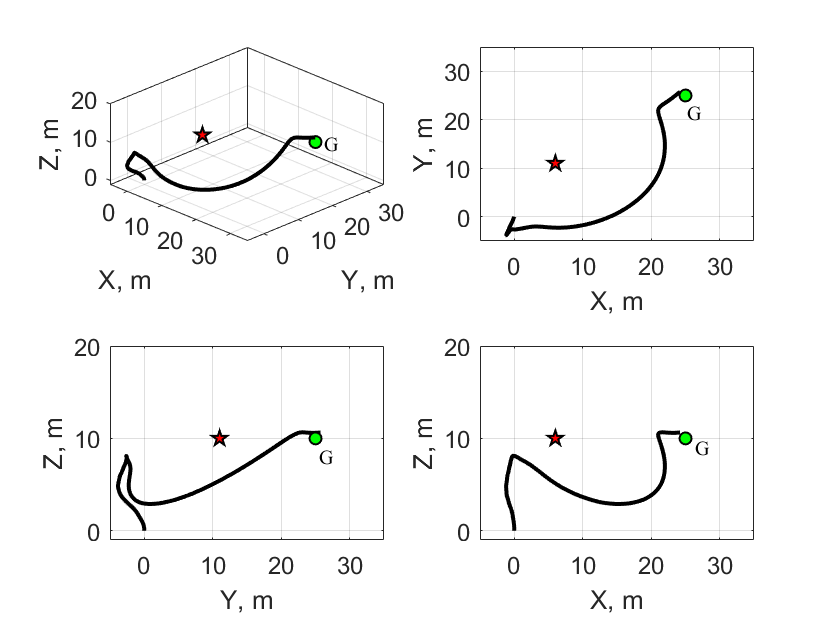}
    \caption{Payload trajectory with obstacle avoidance}
    \label{fig:19}
\end{figure}

\begin{figure}
    \centering
    \includegraphics[width=\columnwidth]{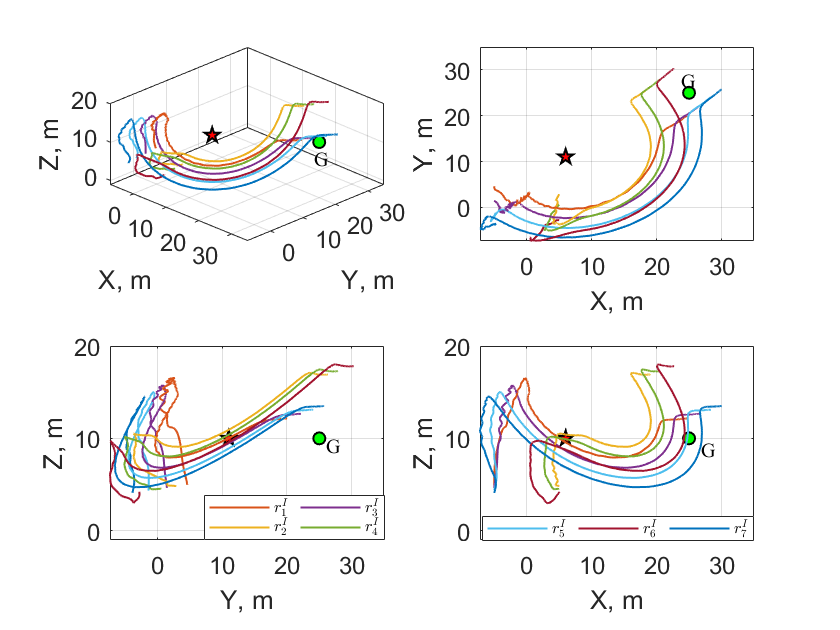}
    \caption{Swarm trajectory with obstacle avoidance}
    \label{fig:20}
\end{figure}

\begin{figure}
    \centering
    \includegraphics[width=\columnwidth]{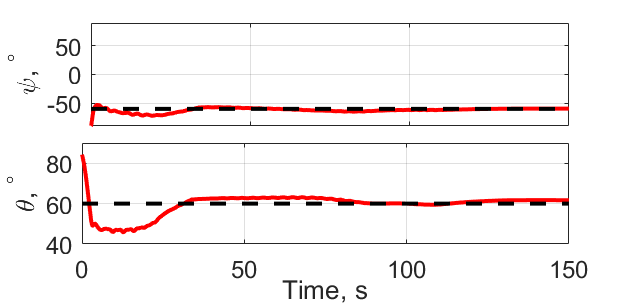}
    \caption{Azimuth and elevation angle of the payload with obstacle avoidance}
    \label{fig:21}
\end{figure}

\begin{figure}
    \centering
    \includegraphics[width=3in]{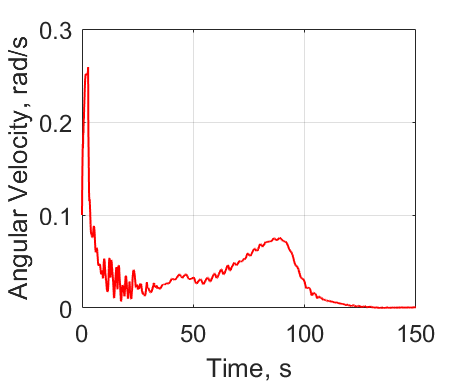}
    \caption{Angular velocity of the payload with obstacle avoidance}
    \label{fig:22}
\end{figure}

An obstacle is located at $[6,11,10]^T$. Figure \ref{fig:19} shows the maneuver deviation of the payload around the obstacle (shown as a star) as the agents detect it. This deviation is the obstacle avoidance behaviour generated from Eq. \eqref{eq:45} in the decentralized swarm control law. Figure \ref{fig:20} shows the corresponding swarm trajectory to move the payload from origin to goal set point. To achieve the given azimuth and elevation angles, the swarm agents maintain formation at different heights using Eq. \eqref{eq:43} as shown in y-z and x-z projections in Fig. \ref{fig:20}. The desired azimuth and elevation angles are achieved for the payload as shown in Fig. \ref{fig:21}. The angular velocity of the payload stays close to zero throughout the mission as shown in Fig. \ref{fig:22}.

\subsection{Case 2: Transportation and manipulation in presence of disturbance}
\begin{figure}
    \centering
    \includegraphics[width=\columnwidth]{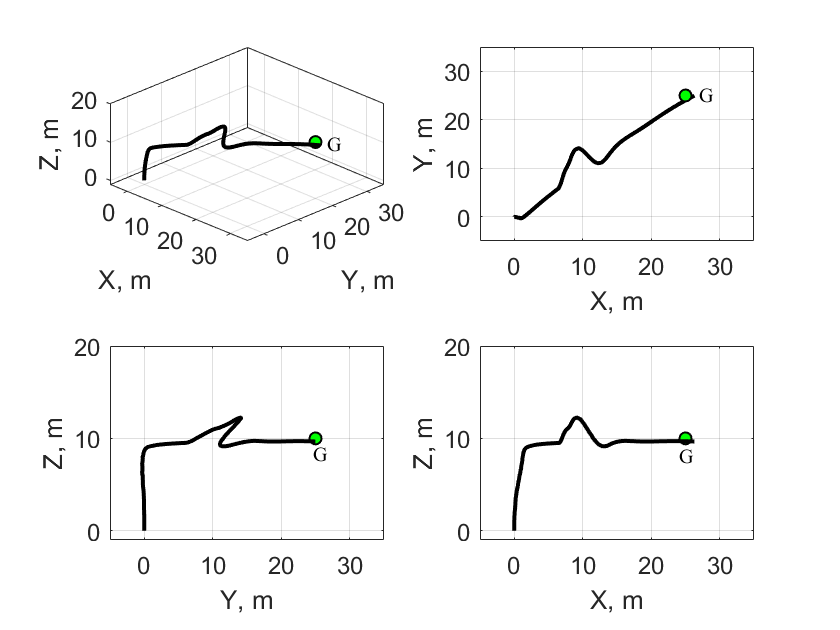}
    \caption{Payload trajectory under wind disturbance}
    \label{fig:11}
\end{figure}

\begin{figure}
    \centering
    \includegraphics[width=\columnwidth]{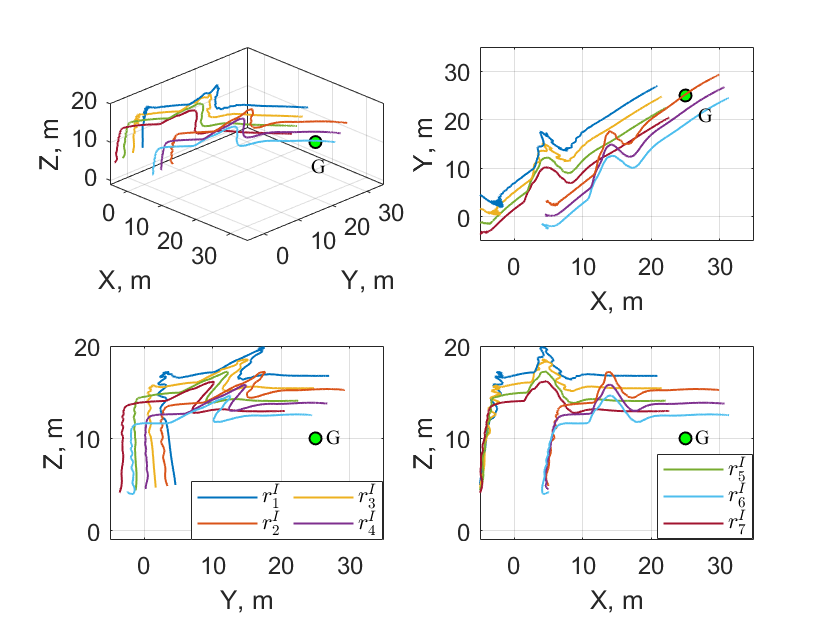}
    \caption{Swarm trajectory under wind disturbance}
    \label{fig:12}
\end{figure}

\begin{figure}
    \centering
    \includegraphics[width=\columnwidth]{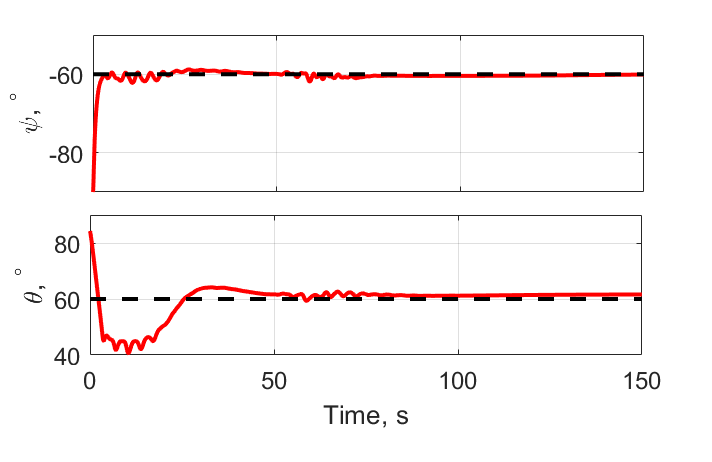}
    \caption{Azimuth and elevation angle of the payload under wind disturbance}
    \label{fig:13}
\end{figure}

\begin{figure}
    \centering
    \includegraphics[width=2in]{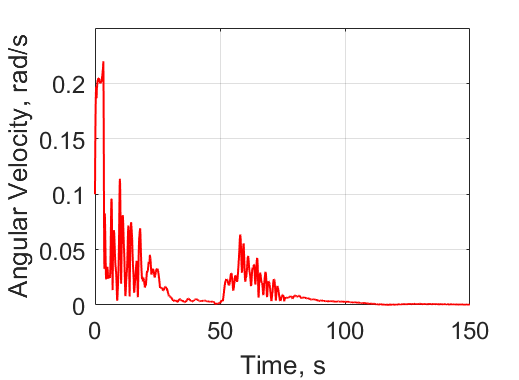}
    \caption{Angular velocity of the payload under wind disturbance}
    \label{fig:14}
\end{figure}

Figure \ref{fig:11} shows the trajectory of the payload. The x-y, y-z, and x-z projections of the trajectory are also shown in the subplots. The corresponding swarm trajectories in different views are shown in Fig. \ref{fig:12}. The azimuth and elevation angles of the payload are shown in Fig. \ref{fig:13}. Figure \ref{fig:14} shows the angular velocity of the payload during the maneuver. An oscillating force is introduced in the system (between $t = 50s$ to $t = 60s$) to replicate wind disturbance on the payload. The trajectories in Fig. \ref{fig:11} show that these disturbances are attenuated by the swarm system. The angular velocity of the payload also damps to zero as shown in Fig. \ref{fig:14}.

\subsection{Case 3: Transportation and manipulation in case of failure}
\begin{figure}
    \centering
    \includegraphics[width=\columnwidth]{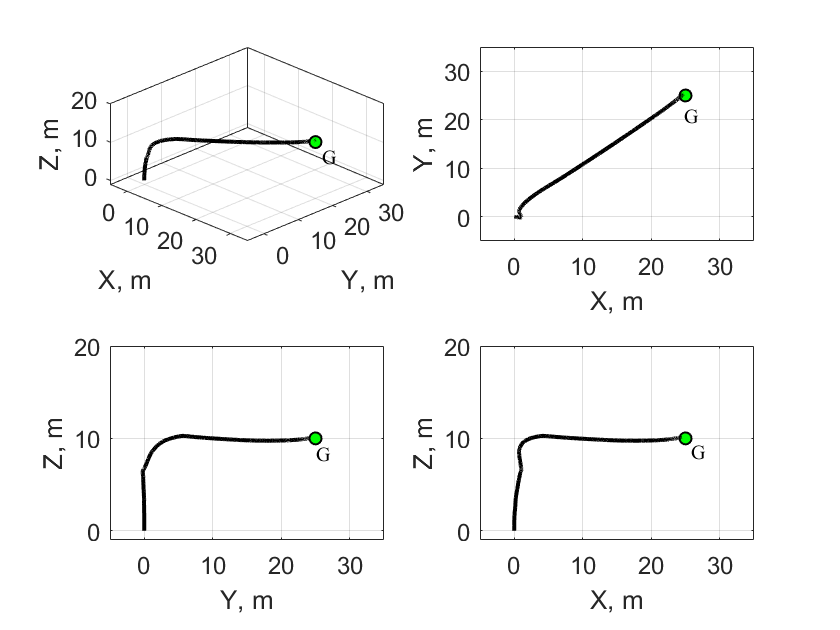}
    \caption{Payload trajectory under agent failure}
    \label{fig:15}
\end{figure}

\begin{figure}
    \centering
    \includegraphics[width=\columnwidth]{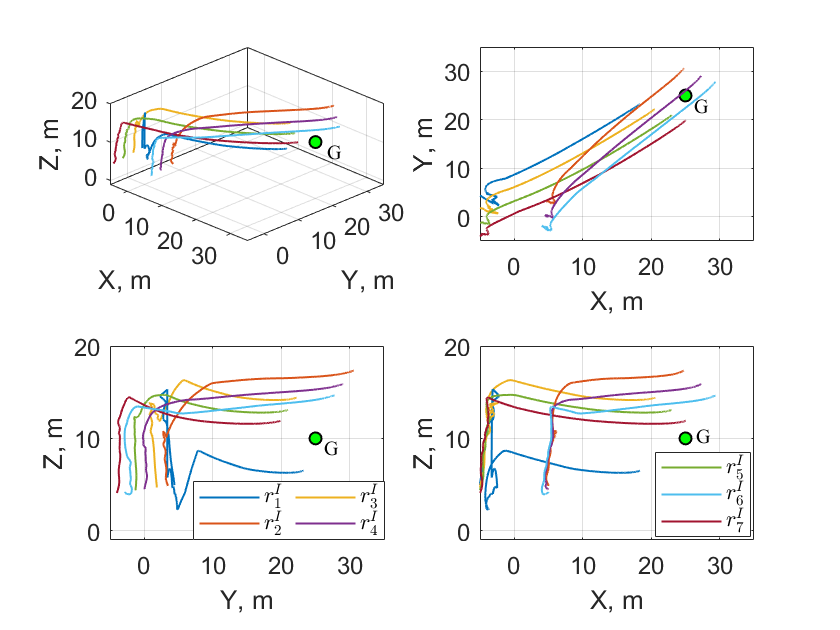}
    \caption{Swarm trajectory under agent failure}
    \label{fig:16}
\end{figure}

\begin{figure}
    \centering
    \includegraphics[width=\columnwidth]{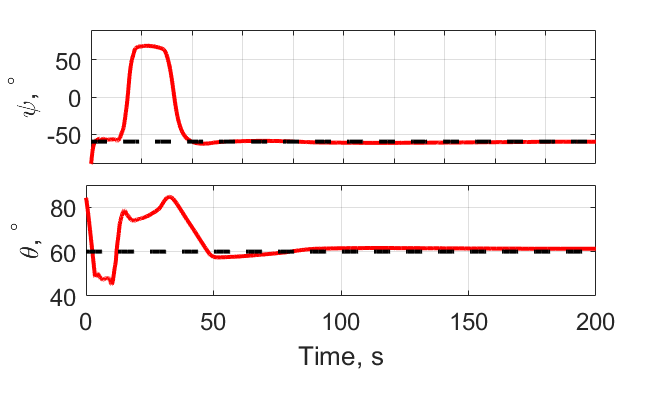}
    \caption{Azimuth and elevation angle of the payload under agent failure}
    \label{fig:17}
\end{figure}

\begin{figure}
    \centering
    \includegraphics[width=2in]{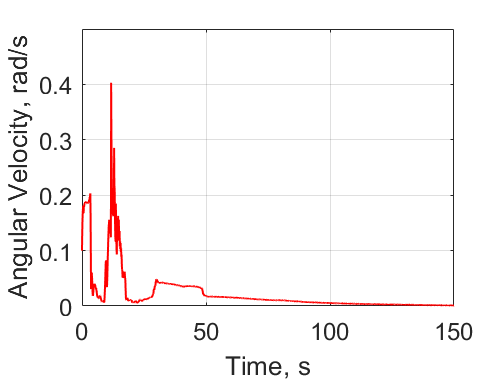}
    \caption{Angular velocity of the payload under agent failure}
    \label{fig:18}
\end{figure}

A decentralized control scheme has the advantage of resilience to failure, other than scalability and flexibility. A loss of control on an agent is simulated at $t \geq 10s$. The payload trajectory is not much affected by this loss of control as shown in Fig. \ref{fig:15}. In Fig. \ref{fig:16}, the loss of control on swarm agent with index 1 can be seen as trajectories of agent $1$ drops instantly and hangs to the payload. A sudden degradation of azimuth and elevation angle tracking is observed near the point of failure ($t=10s$) as shown in Fig. \ref{fig:17}. A spike in angular velocity of the payload is observed at the point of failure in Fig. \ref{fig:18}. Results from Fig. \ref{fig:17} and \ref{fig:18} show that such a decentralized approach is capable of carrying out a mission with desired payload transporation and manipulation setpoints even in the case of failure of an agent in the swarm. Moreover, the swarm also has no information about this failure; but the \textit{emergent behaviour} from local interactions of the system tends to compensate for loss of an agent.

\section{Conclusions}
In this paper, a novel approach for payload transportation and manipulation is discussed using a swarm of agents tethered physically to the payload. A decentralized formation of swarm is proposed based on Artificial Potential Field approach for this task. The dynamics of the swarm and the payload are studied considering the cable as a series of lumped masses connected \textit{via} spring-damper systems. An equilibrium study for this nonlinear-coupled system is performed and the stability of this equilibrium is proved. The proposed swarm formation is able to manipulate the payload's azimuth and elevation angles. A novel transportation field is introduced to track a virtual swarm center for transportation of the payload without information of the payload states. This control signal is passed through a PID block to compensate for asymmetric swarm anchor points, irregular mass distribution of payload, and other unaccounted effects (wind, failure of agents etc.). The proposed method is verified \textit{via} numerical simulations in three cases: performance in presence of obstacle, wind disturbance, and failure of an agent. The swarm successfully completes the mission in all the three cases, highlighting the resilience and robustness of such decentralized scheme.

\section{Conflict of Interest}
All authors declare that they have no conflicts of interest.

\bibliography{IEEEabrv,sample}
\bibliographystyle{IEEEtran}

\end{document}